\documentclass{article}
\usepackage{spconf,amsmath,epsfig}
\usepackage{algorithm}
\usepackage{booktabs} 
\usepackage{epsfig}
\usepackage{amssymb}
\usepackage{amsmath}
\usepackage{tabularx}
\usepackage{tabularx}
\usepackage{tabularx,ragged2e}
\newcolumntype{C}{>{\Centering\arraybackslash}X} 
\usepackage{amsfonts}
\usepackage[utf8]{inputenc}
\usepackage[T1]{fontenc}
\usepackage{tabu}
\usepackage{hyperref}
\usepackage{multirow}
\usepackage{makecell}
\usepackage{ctable}
\usepackage{capt-of}
\usepackage{varwidth}
\usepackage{booktabs} 
\usepackage{amssymb}
\usepackage{graphicx}
\usepackage{algorithm}
\usepackage{algpseudocode}
\usepackage{pifont}
\usepackage{url}
\usepackage{caption}

\begin{document}\sloppy

\def\x{{\mathbf x}}
\def\L{{\cal L}}

\title{Endogenous and Exogenous Multi-Modal Layers in
Context Aware Recommendation Systems for Health}
%
\name{Nitish Nag, Vaibhav Pandey, Ramesh Jain}
\address{Department of Computer Science, University of California, Irvine - California, USA}
\maketitle

\begin{abstract}
People care more about the solutions to their problems rather than data alone. Inherently, this means using data to generate a list of recommendations for a given situation. The rapid growth of multi-modal wearables and sensors have not made this jump effectively in the domain of health. Modern user content consumption and decision making in both cyber (e.g. entertainment, news) and physical (eg. food, shopping) spaces rely heavily on targeted personalized recommender systems. The utility function is the primary ranking method to predict what a given person would explicitly prefer. In this work we describe two unique layers of user and context modeling that can be coupled to traditional recommender system approaches. The exogenous layer incorporates factors outside of the person's body (eg. location, weather, social context), while the endogenous layer integrates data to estimate the physiologic or innate needs of the user. This is accomplished through multi-modal sensor data integration applied to domain-specific utility functions, filters and re-ranking weights. We showcase this concept through a nutrition guidance system focused on controlling sodium intake at a personalized level, dramatically improving upon the fixed recommendations.
\end{abstract}
\section{Introduction}
Recommendation systems are becoming an increasingly important part in various domains of daily life. The large sets from which users must make decisions from can be overwhelming from too much choice. In areas such as online shopping, entertainment, and information search there has been considerable effort to streamline choice through recommendations. These traditional recommendation systems consider two main components, the user and the items. In more recent work, context has become the third valuable component. Context is very complex and difficult to characterize, but usually has an affect to alter the recommendations to fit a given user situation. Building real-time user models from context has been a key component of advanced recommender systems. Generally, the contextual information of a given user is divided into 3 classes of fully observable, partially observable, or unobservable \cite{Adomavicius2015Context-AwareSystems}. Incoming data to determine context is divided into another dimension of static or dynamic data. Wearable devices, sensors, mutli-media, social networks, and many more multi-modal data streams carry information that can better characterize the user model to enhance recommendations in both static and dynamic data domains. Applying these multi-modal data sources for health applications remains a challenge. The growth momentum of wearable devices that monitor health metrics via sensors continues to promise better health for users. How will the data generated by these sensors become useful in daily life? Ideally an important application would be to improve personalized recommendations pertaining to health decisions. In this work we extend a new dimension to contextual data by separating the distinction between information in the environmental situation of a user (exogenous to the user) and the inherent biological needs of a user under the skin (endogenous to a user). After an explanation of the concept, we show this in real-time prediction of sodium needs for food recommendations.

\begin{figure}
\small
\centering
\includegraphics[width=0.90 \columnwidth]{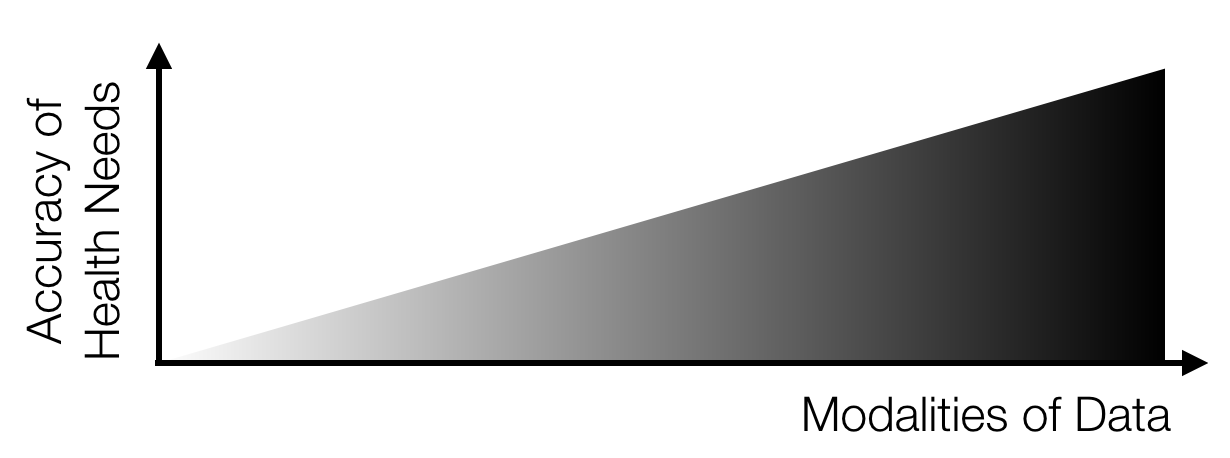}
\caption{Increasing aggregation of mutli-modal user data streams can more accurately predict the health status and needs of a user.}
~\label{fig:needs}
\vspace{-5mm}
\end{figure}

\section{Related Work}
Context-Aware Recommendation Systems (CARS) in various domains have been used to improve recommendation performance increasingly over the past decade. Basic techniques of pre/post-filtering and context modeling have been well described in reviews and texts \cite{Adomavicius2015Context-AwareSystems}. Furthermore, the contextual data has been usually split into dynamic and static components that can be observed at various fidelity levels (X and Y axis on Figure \ref{fig:endo}). 

Individual context is a major category of CARS as described by Villegas et. al. \cite{Villegas2018CharacterizingReview}. Within individual context, Villegas describes a sub category of "natural" events which are characteristics human intervention does not control. This is in reference to weather or pollution. The "human" subcategory describes preferences and behavior. One critical aspect these classifications do not address is the natural biological / health context of the individual, which is the main goal of this work.

Using multimedia and multi-modal data to power health recommendations has been echoed as a need in previous work by several groups in the multimedia and recommender communities \cite{Nag2017} \cite{Schafer2017TowardsSystems}. A big part of this is understanding what the user needs through either direct understanding of intent \cite{Kofler2016UserSearch} or through implicit understanding of user needs, especially in the area of health from sensors and wearable devices \cite{Lu2015RecommenderSurvey}. Health recommender systems have two main user groups that are well described: The patient and the health-care provider \cite{Wiesner2014HealthChallenges} \cite{Sezgin2013ASystems}. Both of these user groups need to know biological context to apply recommendations that affect health.

\section{Biological Context: Exogenous and Endogenous Layers}

\begin{figure}
\small
\centering
\includegraphics[width=0.92 \columnwidth]{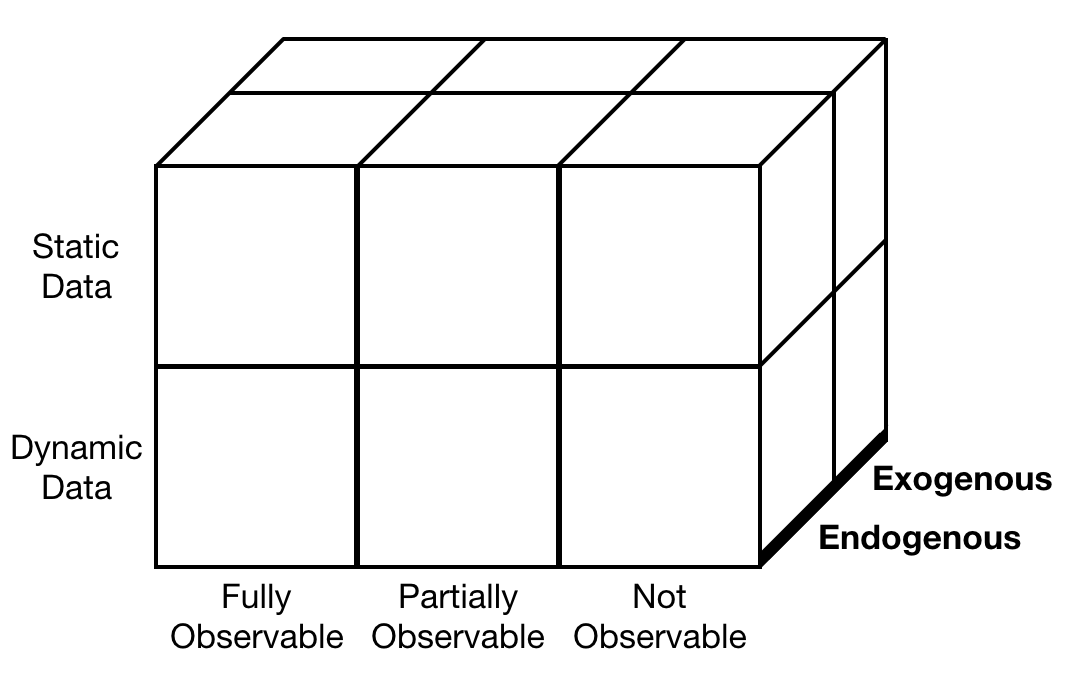}
\caption{The traditional context data dimensions of data change rates and context knowledge compose the X and Y axis. In health we expand the Z axis dimension with exogenous and endogenous layers. An example to consider would be glucose monitoring: A single finger prick reading would be a partially observable endogenous context data type, whereas a continuous glucose monitoring device would be a fully observable endogenous context data type. Location based services would fall in the exogenous layer.}
~\label{fig:endo}
\vspace{-5mm}
\end{figure}

Recommendation in health is inherently different from the popular recommendation systems most users interact with that try to estimate user preference or ratings (Eg. Netflix, Amazon etc.). We need to consider at least two aspects while recommending an item to a user in the context of health. 1) \textit{How item affects the users health}, 2) \textit{How does the item align with user's preferences (traditional recommender approach)}. The Endogenous and Exogenous Layers in this work considers this first component of health in an organized fashion. The same item may affect a user in a different way given different situations. The layers in this concept are clearly delineated between the environmental factors external to the skin and physiologic factors beneath the skin. Combined, this gives a clean separation of situations in the biological body in order to use physiological science to predict needs (Figure \ref{fig:endo}).

\subsection{Exogenous Context}
Exogenous context defines the realm of the situation outside the body. The environmental factors that can be determined include the physical, social, and information external to the user. The sensors providing the data about this may not come from a user device, and from stations in the environment that are monitoring the aforementioned components. For example, pollution monitors, weather stations, traffic data and events in the external space are monitored by specialized fixed sensors. Using the user location from the device we can derive which sensors to pull data from to estimate the user environment. Some considerations to take are if the user is inside or outside a building. The GPS location may give a general area, but if the user is inside the building, we must know the environmental status inside the building, not the outside weather. Sensors on the user may also capture information about the environment since most wearables and smartphone used to capture data are external to the user body.

\subsection{Endogenous Context}
Endogenous context defines the physiological status of the user physical body. Many sensors and wearable devices are focused on capturing this data. As an example, heart rate monitors, thermometers, glucose monitors, blood pressure readings, accelerometers for movement, skin galvanic resistivity sensors just to name a few. For sport specific and military applications this is further expanded with embedded sensors, power meters, and mutli-location accelerometers on the body. These sensors are all trying to estimate some physiological parameter of the body. Integration of these sensors into a meaningful "body / health context" for use in recommender systems is essential for these sensors to become meaningful for daily life \cite{Nitish2018Cross-ModalEstimation}. The context will be used in domain specific applications determined by the use case by the user or health provider. For example, in diabetes, glucose monitoring along with the other sensors has a goal of reducing the global total plasma glucose levels to control the disease. Yet these sensors are still unable to provide common daily life recommendations for diabetes management like what a user should be eating. In the case of heart failure (most expensive readmission hospital cost in the United States \cite{Hines2011Conditions2011}), blood pressure monitoring along with other factors does not control the primary risk factor in readmission to the hospital. A primary driver of readmission remains high sodium food intake, which results in fluid retention and exacerbation of the heart failure. Endogenous context in this case could be used to make recommendations to the patient to eat foods that have the appropriate level of sodium in the food. This is the specific aim we tackle in the application of this concept.

\begin{figure}
\small
\centering
\includegraphics[width=0.85 \columnwidth]{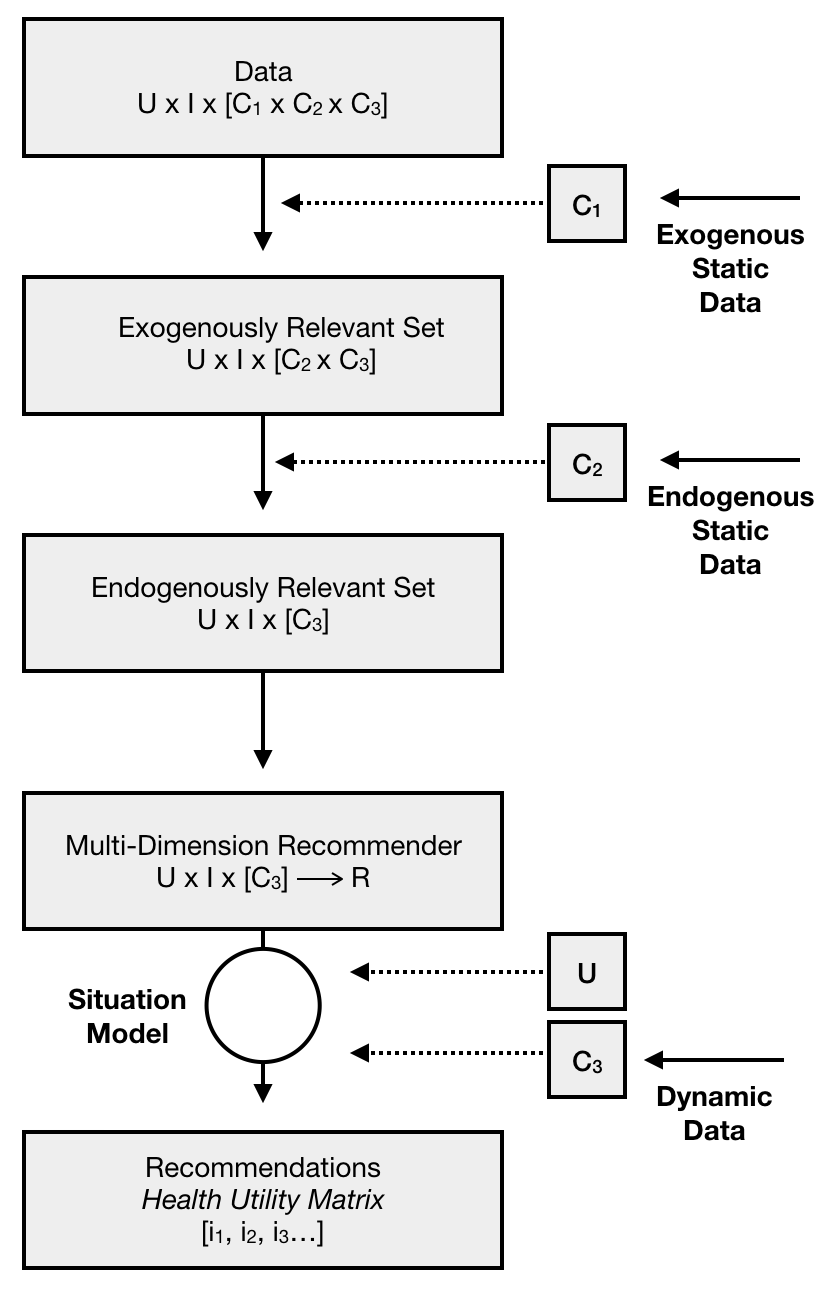}
\caption{A hybrid of two paradigms used in CARS. Pre-filtering and multi-dimensional modeling can be an effective method for health contextual understanding. The relevant items are reduced in space through the pre-filter of exogenous and endogenous static data. This reduced set can then be incorporated into a multi-dimensional model that takes into account dynamic data streams from multi-modal data to give the final utility matrix of items.}
~\label{fig:flow}
\vspace{-5mm}
\end{figure}

\subsection{Integration of Biological Context Layers in CARS}
A method of bringing these layers into CARS is shown in Figure \ref{fig:flow}. The initial work flow uses the CARS tools of pre-filtering with static data parameters to reduce the item space only to relevant and safe to consume items. Following this we use a multi-dimension approach to integrate both the exogenous and endogenous dynamic data. The reason we must combine these two layers in this phase is because exogenous factors affect the endogenous state. In the case of sodium needs, the temperature and humidity will affect the sweat rate of an individual. This situation modeling portion must be customized to the domain specific use case and goals of the recommender system. The end goal of this work flow is not to have a guessed preference score like in traditional recommender systems (hence no reference of Ratings in the data space), rather to estimate the user health needs.

\section{Application to Health: Nutrition Recommendations}
\subsection{System Goals}
To capture the user attributes necessary to elucidate the exogenous and endogenous context, we create a multi-modal user vector using different sources of data such as smartphones and sensors (body weight, accelerometer based motion, barometric altitude gain/loss, timestamps) and environmental data (altitude, temperature). We need to incorporate expert knowledge in the system to bring the user vector and item vector in the same space in order to compute the utility matrix. We are doing that using a combination of various algorithms in Table \ref{tab:sodiumscience}, to convert the relevant aspects of the user vector to the space of the item vectors (i.e. nutritional sodium requirements) and matching them to generate a score for how well the corresponding item satisfies the user's sodium nutritional needs \cite{Nag2017PocketLocation}. As mentioned earlier, this is a great need for heart failure patients to have tailored sodium intake, yet the gold standard is a single uniform number (1500mg) given by the American Heart Association \cite{Gupta2012DietaryFailure}.

\begin{table}[t]
\begin{center}
\caption{Scientific Resources to Understand Sodium Needs} \label{tab:sodiumscience}
\begin{tabular}{|c|c|}
  \hline
  Component & Author
  \\
  \hline
  Temperature & Bates et. al. \cite{Bates2008SweatHeat}  \\
  Altitude & Hannon et. al. \cite{Hannon1971AlterationsExposure.}  \\
  Walking & Howley et. al. \cite{Howley1974TheWomen.}  \\
  Stairs & Aziz et. al. \cite{Aziz2005PhysiologicalStairs}  \\
  Basal Metabolism & Schofield et. al. \cite{Schofield1985PredictingWork.}  \\
  Age & Stapleton et. al. \cite{Stapleton2014Age-relatedPostexercise}  \\
\hline
\end{tabular}
\end{center}
\end{table}
\vspace{0mm}

\begin{figure}
\small
\centering
\includegraphics[width=1.00 \columnwidth]{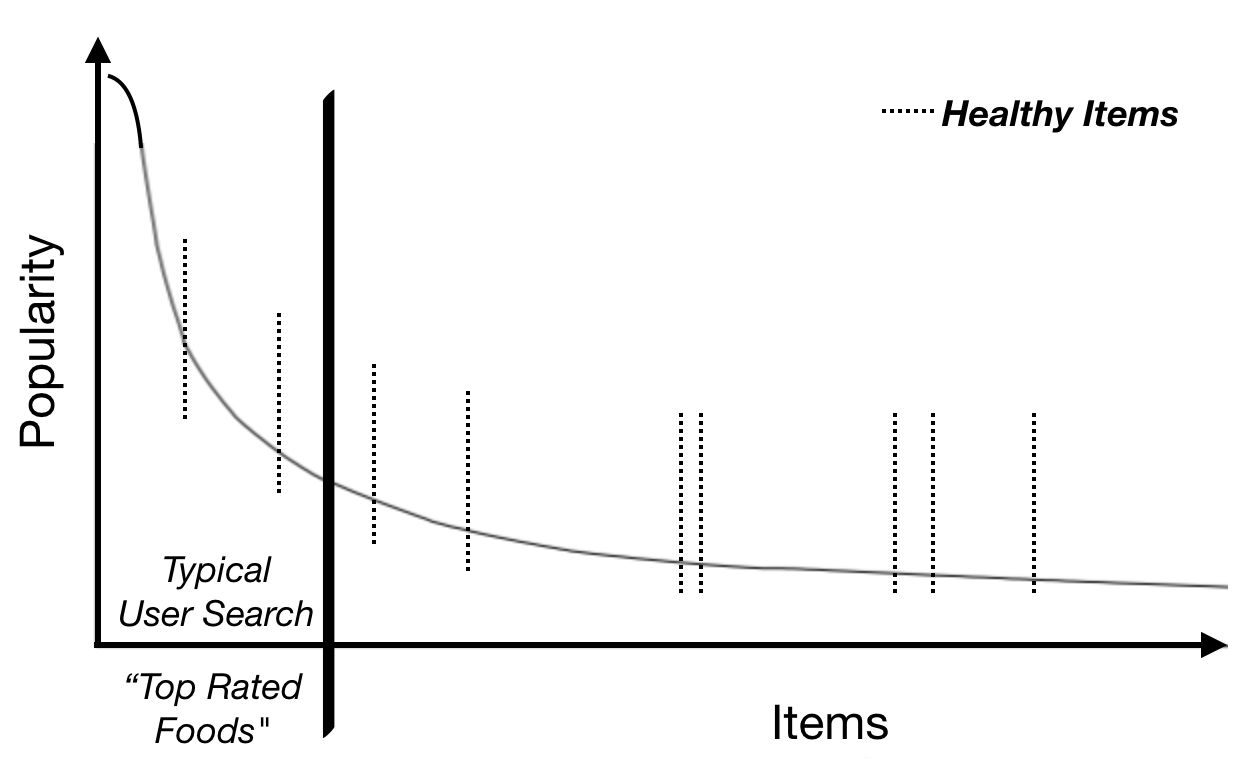}
\caption{In consumer decisions, the long tail problem makes it difficult to find items that are not popular but may be relevant to the user. By understanding the health status of the user, CARS can match health relevant items that a user would not usually browse through. This essentially allows the user to discover healthy items that are relevant but would have been difficult to find through popularity based rating systems.}
~\label{fig:tail}
\vspace{-5mm}
\end{figure}

This is a real-world problem as users use electronic review sources very often to look for food. These services (e.g. Yelp, Zomato, OpenTable etc.) do not take any health factors into consideration. The big problem for users when deciding on where to eat is that they search on the granularity level of the restaurant. This is not the appropriate way to match a user to a meal, as the restaurant may have various offerings that are healthy or unhealthy. The correct way to match a user to a meal would be to match at the item level of the meal. This is parallel to shopping online for a store versus an item. The problem with the online resources for food is that they are so large, a user cannot browse through all the options. Thus a recommender system should provide relevantly matched items to reduce browsing burden. This problem is commonly known as the Long Tail Problem (Figure \ref{fig:tail}). The goal of this system is to capture the healthy items in the long tail and match them to the personal needs of a user based on sodium needs.

\subsection{System Architecture}
\begin{figure}
\small
\centering
\includegraphics[width=1.00 \columnwidth]{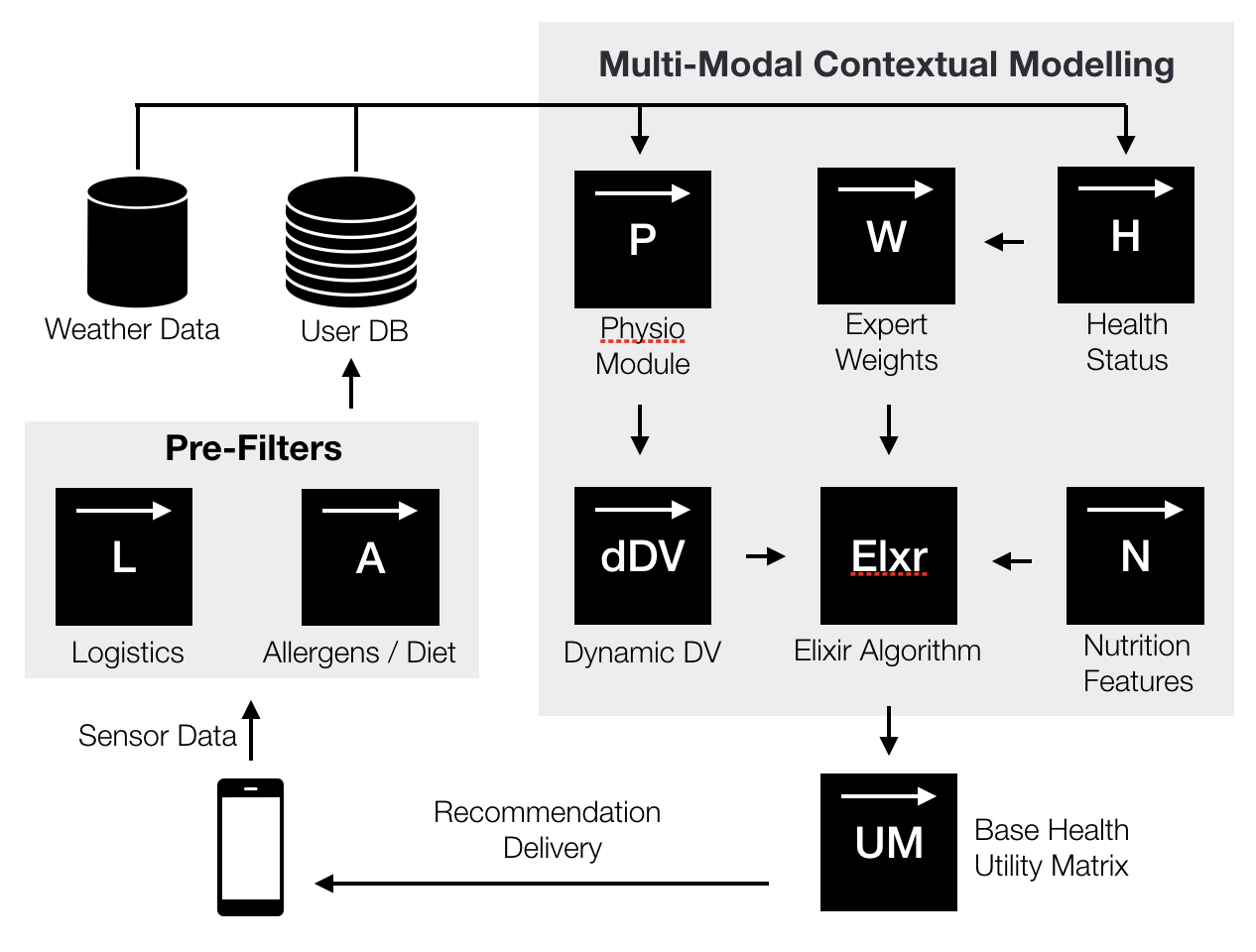}
\caption{The conceptual architecture of the nutrition recommendation system. The Physio Module estimates the user physiological response to the endogenous and exogenous factors. These are mapped to the same dimensions as the nutrition features (sodium in this case) to allow for combination using the Elixir algorithm to rank foods \cite{Nag2017PocketLocation}.}
~\label{fig:arch}
\vspace{-5mm}
\end{figure}

We adapt the paradigms of pre-filtering and multi-dimensional analysis from Figure \ref{fig:flow} for the meal recommendation problem in Figure \ref{fig:arch}. This system uses the smartphone as a sensor capture device and for the recommendation delivery. 

In reference to the work flow described in Figure \ref{fig:flow}, we use a database crawled from publicly available restaurant menus on the internet to gather nutrition facts and restaurant information to generate the item space. The exogenous pre-filter takes into account static data pertaining to availability of meal resources including timing, distance away and traffic conditions. The endogenous pre-filter takes into account static data pertaining to the users food allergens and dietary preferences (vegan, vegetarian). The Multi-Dimensional situation model uses an algorithm (below) derived from biological scientific studies in measuring sodium loss and retention in various situations as described in Table \ref{tab:sodiumscience}. 

\subsection{Experiments and Results}
The primary aim of our application system is to find the healthiest (defined by correct sodium needs) meal for purchase given a location. We have three different synthetic scenarios (Table \ref{tab:scenario}) that we test three artificial users (Table \ref{tab:users}). After identifying the sodium loss from calculations below derived from literation, we get a sodium need per user-context situation (Table \ref{tab:sodium}), we try to match this to the most appropriate meal within a 30km radius. The ranking of the foods is then carried out by the food ranking algorithm Elixir \cite{Nag2017PocketLocation} on the subset of pre-filtered items.

\begin{eqnarray}
kmWalked &=& (0.762)Steps/1000  \nonumber \\
StepCalories &=& (0.67)(kmWalked)(Weight) \nonumber \\
StairCalories &=& (0.026)(Floors/3)(Weight) \nonumber \\
BMR &=& see \cite{Schofield1985PredictingWork.} \nonumber \\
DailyCalories &=& BMR + StairCalories + StepCalories \nonumber \\
BasicNa &=& (1.2)DailyCalories \nonumber \\
TempNa  &=& (.74)Celsius \nonumber \\
AltiNa  &=& (meters/300)^2.5 \nonumber \\
TotalSodium  &=& BasicNa + TempNa + AltiNa \nonumber \\
\end{eqnarray}

\begin{table}
\small
\centering
\caption{Synthetic Users. Health Status: N = Normal, O=Obese, MA=Muscle Atrophy}
\label{tab:users}
\begin{tabular}{@{}llclllll@{}} \Xhline{3\arrayrulewidth}
&                       & \multicolumn{1}{l}{}           & \multicolumn{4}{c}{Health Parameters} \\
 &   & Height & $Weight$  & $Gender$  & $Health$           & $Age$             \\ \hline
\multirow{3}{*}{\rotatebox{90}{Users}} 
& $U_1$ \vline& 167 cm       & 125 lbs & Male & N          & 29                 \\
& $U_2$ \vline& 190 cm       & 290 lbs& Male & O          & 37                   \\
& $U_3$ \vline& 155 cm        & 85 lbs   & Female  & MA           & 18            \\
\Xhline{3\arrayrulewidth}
\end{tabular}
\vspace{-2mm}
\end{table}

\begin{table}
\small
\centering
\caption{User scenarios. For generating latitude/longitude we use the following: Workday in Los Angeles, California inside office Building. Hiking in Yosemite National Park, USA. Beach refers to Newport Beach, California.}
\label{tab:scenario}
\begin{tabular}{lllll}
\Xhline{3\arrayrulewidth}
\multirow{2}{*}{Situation} & \multicolumn{4}{c}{Sensor Parameters}                                                  \\ \cline{2-5} 
                           & \multicolumn{1}{c}{Steps}  & \multicolumn{1}{c}{Floors} & \multicolumn{1}{c}{Altitude} & \multicolumn{1}{c}{Temp} \\ \hline
Workday &  \multicolumn{1}{c}{2,400} &  \multicolumn{1}{c}{12} &  \multicolumn{1}{c}{20 feet} & \multicolumn{1}{c}{70 F}     \\ 
Hiking        &  \multicolumn{1}{c}{30,650} &  \multicolumn{1}{c}{207} &  \multicolumn{1}{c}{10,700 feet} & \multicolumn{1}{c}{42 F}      \\
Beach Picnic           &  \multicolumn{1}{c}{7,430} &  \multicolumn{1}{c}{31} &  \multicolumn{1}{c}{0 feet} & \multicolumn{1}{c}{92 F}   \\ \Xhline{3\arrayrulewidth}                  
\end{tabular}
\end{table}

\begin{table}
\caption{Sodium Need Results}
\label{tab:sodium}
\begin{tabular}{@{}llclllll@{}} \Xhline{4\arrayrulewidth}
&                       & \multicolumn{1}{l}{}           & \multicolumn{5}{c}{Health Parameters} \\
 &   & Siutation & $Calories/Day$  & $Sodium (mg)$    \\ \hline
\multirow{11}{*}{\rotatebox{90}{Users}} 
& $U_1$ \vline& Workday       & 2767 & 3320      \\
& $U_1$ \vline& Hiking       & 3798& 4558        \\
& $U_1$ \vline& Beach        & 2826   & 3391          \\
& $U_2$ \vline& Workday     & 3712 & 4455     \\
& $U_2$ \vline& Hiking       & 6022   & 7277         \\
& $U_2$ \vline& Beach        & 3923   & 4708              \\
& $U_3$ \vline& Workday     & 2183 & 2620      \\
& $U_3$ \vline& Hiking       & 2894   & 3472          \\
& $U_3$ \vline& Beach        & 2215   & 2658                           \\
\Xhline{\arrayrulewidth}
\end{tabular}
\end{table}

\section{Future Work and Conclusions}
While we have demonstrated the system with only a health utility function for the healthy meal recommendation problem, this approach could be generalized to include other utility functions and CARS techniques which capture different aspects of the user's decision making process affecting health. Future work to advance health needs profiling of a user will need to take into account many factors including the following:
\subsection{Behavioral and Social Understanding}
Humans are creatures of habit. Understanding the behavioral reasons for decisions made using recommender systems is critical to effectively engage the user in choosing the healthiest choices from a health recommender system. Behavioral modification remains a complex and difficult task in health applications. For example, a user may exhibit different behavior when in presence of friends and family due to peer pressure which can be captured by identifying their choices in presence of different people. This approach can also be adapted to solve problems in other domains, we would need to replace the expert knowledge module with a relevant expert system or a learning system which could model the behavioral utility function.
\subsection{Multi-Criteria}
Once we have computed the utility matrices capturing different aspects of the user's decision process, we need to combine them to obtain better real-world final recommendations. This can be done by treating health and traditional rating systems as different recommender systems producing their own utility functions. Combining these utility functions can be accomplished in different ways, such as using a weighted average of different utility matrices where the weight of each matrix captures the extent to which the user exhibits the behavior aspect captured by the utility function, or using a threshold based system and chaining the output of one system to another system's input.
\subsection{Cross-Domain Integrations}
There are multiple avenues where recommender systems can apply for users. Knowledge from various domains may need to be combined in the case of certain health situations. Furthermore, knowledge from a source domain may have implications for a different target domain. There are some established approaches for general cross-domain recommender in health. Applying these to the endogenous and exogenous layers remains to be explored. Two broad categories for this exist: 1. Multi-domain: Approaches may be used to understand the way various health source domain factors interact to various target recommendations. In the endogenous setting, this is closely mimicked through the way various organ systems interact and affect each other. For example, readings from a glucose monitor, heart rate monitor, and blood pressure monitor all are in different domains, but may all affect recommendations that result in changes in all domains. 2. Linked and Cross Domains: Knowledge about one domain may inform the best recommendation in a different domain. For example, atmospheric exogenous domain may inform what the user would find most tasty (hot weather may increase the likelihood of saltier food increasing in utility in both the health and user preference matrix).

In conclusion, this work introduces the concept of extending traditional CARS approaches in health through the dimension of exogenous and endogenous components. This distinction may be a critical step organizing the integration of wearable and multi-modal data into recommender systems. This layering allows mapping of the needs for both consumer and health professional situations. These two components also depart from the traditional approach of estimating a user rating, which is a subjective measure of user preference, to an objective measure of user needs. We demonstrate this in a system that is useful for heart failure patients who need customized sodium guidance. Ultimately recommender systems for health are best utilized when they can fulfill these user needs most accurately.

\bibliographystyle{IEEEbib}
\bibliography{Recsys}

\end{document}